\documentclass[11pt]{article}
\usepackage{vietnam,graphicx,multicol}

\newcommand{\dzero}     {D\O}
\newcommand{\rar}       {\rightarrow}
\newcommand{\bbar}      {\mbox{$\bar{b}$}}
\newcommand{\ttbar}     {\mbox{$t\bar{t}$}}
\newcommand{\bbbar}     {\mbox{$b\bar{b}$}}
\newcommand{\qqbar}     {\mbox{$q\bar{q}$}}
\newcommand{\ppbar}     {\mbox{$p\bar{p}$}}

\begin{document}

\begin{flushright}
Fermilab-Conf-04/508-E

{\dzero} Note 4703

November 2004
\end{flushright}

\vspace*{1cm}

\title{REVIEW OF RECENT TOP QUARK MEASUREMENTS}

\author{A. P. HEINSON}

\address{Department of Physics, University of California,\\
Riverside, CA 92521-0413, USA}

\maketitle

\abstracts{
At the Tevatron Collider at Fermilab, a large number of top quarks
have been produced in the ongoing run. The CDF and {\dzero}
collaborations have made first measurements of the {\ttbar} cross
section in several decay channels, and have measured the top quark
mass. In addition, they have set new limits on the cross sections for
single top quark production, and have started to measure some of the
properties of the top quark via studies of its decays. This paper
summarizes the status of these measurements and discusses where they
are heading in the next few years. The paper is based on a talk I gave
at the Rencontres du Vietnam in Hanoi, August 2004; the results have
been updated to show the latest values and new measurements.}

%---------------------------------------------------------------------
\section{Top in a Nutshell}

\begin{figure}[!h!tbp]
\begin{center}
\includegraphics[width=0.90\textwidth]{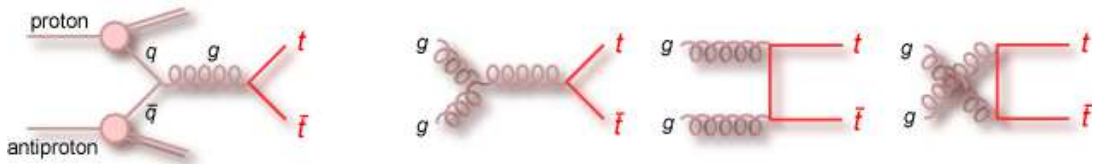}
\end{center}
\caption[fig1]{Tree-level Feynman diagrams for top quark pair
production at the Tevatron Collider.}
\label{feynmanttbar}
\end{figure}

The top quark is a spin +1/2 fermion, with charge +2/3; it is the
weak-isospin partner of the bottom quark, and together the top and
bottom quarks form the third generation of quark families. The top
quark is approximately forty times heavy than its partner, with a mass
of $178.0 \pm 4.3$~GeV measured from the Tevatron Run~I
data.~\cite{topmassrunI} Figure~\ref{feynmanttbar} shows the
tree-level Feynman diagrams for top quark pair production at the
Tevatron. About $85\%$ of the rate comes from the {\qqbar} initial
state and the remaining $15\%$ comes from the $gg$ initial state. The
Standard Model cross section at next-to-next-to-leading order is $6.8
\pm 0.4$~pb.~\cite{topxsectheory}

\begin{table}[!b!h]
\begin{minipage}{13.2cm}
\begin{tabular}{l}
\hline
{\scriptsize Invited talk at the 5th Rencontres du Vietnam, {\it New
Views in Particle Physics,} Hanoi, Vietnam, August 5--11, 2004}
\end{tabular}
\end{minipage}
\end{table}

\clearpage

\subsection{Properties, Production, and Decay}

\begin{figure}[!h!tbp]
\hspace{0.5in}(a)\hspace{2.5in}(b)\vspace{-0.7in}\\
\begin{center}
\hspace{0.3in}
\includegraphics[width=0.30\textwidth]{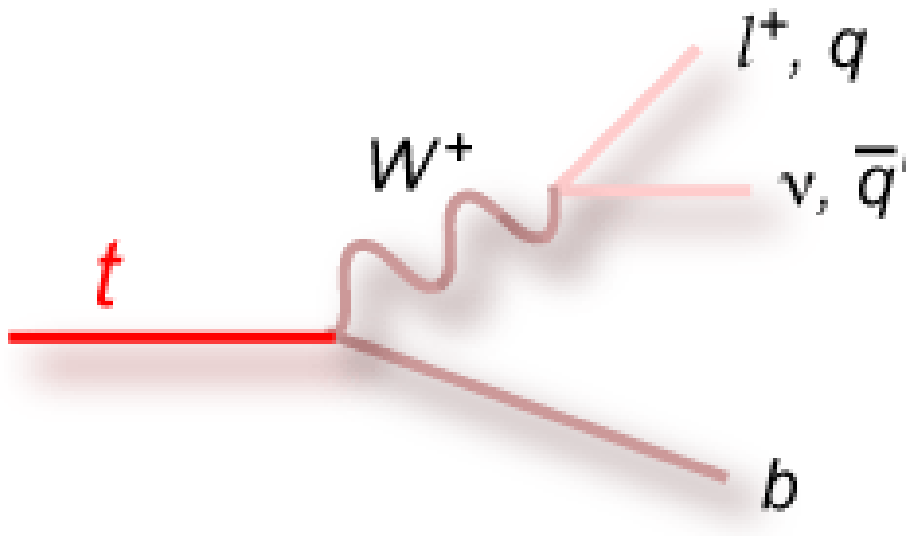}
\hspace{0.8in}
\includegraphics[width=0.40\textwidth]{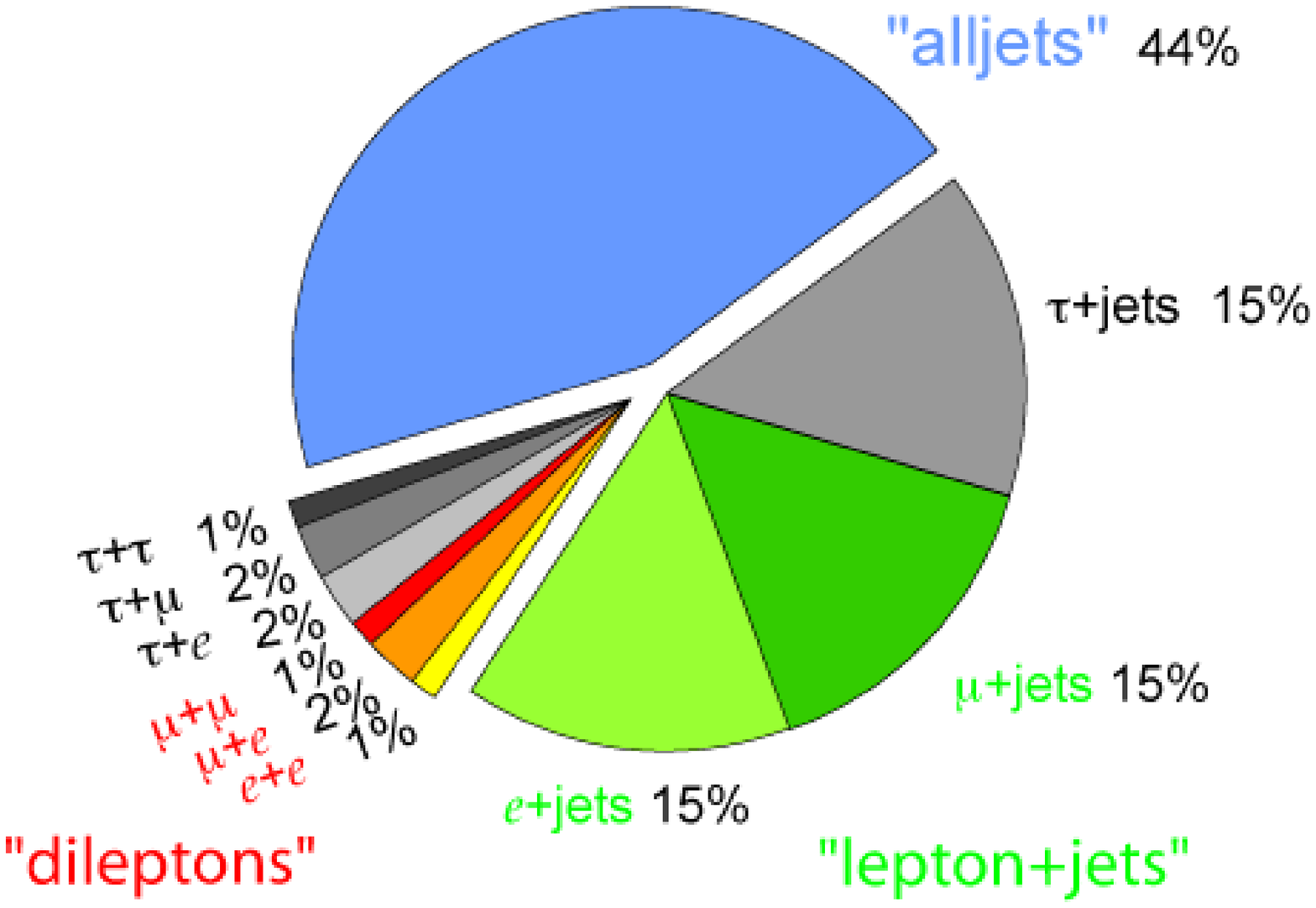}
\hfill
\end{center}
\caption[fig2]{\hspace{0.1in}(a) Tree-level diagram of a top quark
decay.\hspace{0.2in}
(b) Pie chart of the branching fractions for {\ttbar} decays.}
\label{topdecay}
\end{figure}

The top quark decays before it can hadronize, since its lifetime
$\Gamma_{\rm top}^{-1} = (1.5{\rm ~GeV})^{-1}$ is much less than the
QCD scale $\Lambda_{\rm QCD}^{-1} = (200 {\rm ~MeV})^{-1}$. Therefore,
there are no top mesons or baryons. It decays 99.9\% of the time into
$W^+b$, as shown in Fig.~\ref{topdecay}(a). Top pair events are
classified by the decays of the two $W$~bosons: ``dileptons'' ($ee$,
$e\mu$, $\mu\mu$); ``lepton+jets'' ($e$+jets, $\mu$+jets); and
``alljets.'' The branching fractions for these decays are illustrated
in Fig.~\ref{topdecay}(b). The $\tau$ decays are included in the other
channels experimentally, depending on the $\tau$ decay mode.

\vspace{0.5in}

\vspace{-0.1in}
\subsection{Detection and Reconstruction}

Top quarks are produced at the Fermilab Tevatron Collider. This
machine accelerates protons and antiprotons at a center-of-mass energy
of 1.96~TeV, with two collision regions, one at the CDF detector, and
the other at the {\dzero} detector. The collider has been upgraded in
energy and instantaneous luminosity and the detectors have had
significant upgrades to match, including new tracking systems, and
major improvements to the calorimeters and muon systems.

From the first part of Run~II, April 2002 -- August 2004, CDF
collected 370~pb$^{-1}$ of data (330~pb$^{-1}$ with the SVXII silicon
tracker in operation), and {\dzero} collected 470~pb$^{-1}$ of data
(all with the SMT silicon tracker operating). The results shown here
are for the first 100--200~pb$^{-1}$ of data from each collaboration,
several million triggered events. The collaborations select final
samples of events from this data to maximize the measurement
sensitivity, leading to about 100 top quark events above backgrounds
identified so far.

\subsection{Backgrounds}

\begin{figure}[!h!tbp]
\vspace*{0.7in}\hspace{0.5in}(a)\hspace{2.9in}(b)\vspace{-0.9in}\\
\begin{center}
\includegraphics[width=0.25\textwidth]{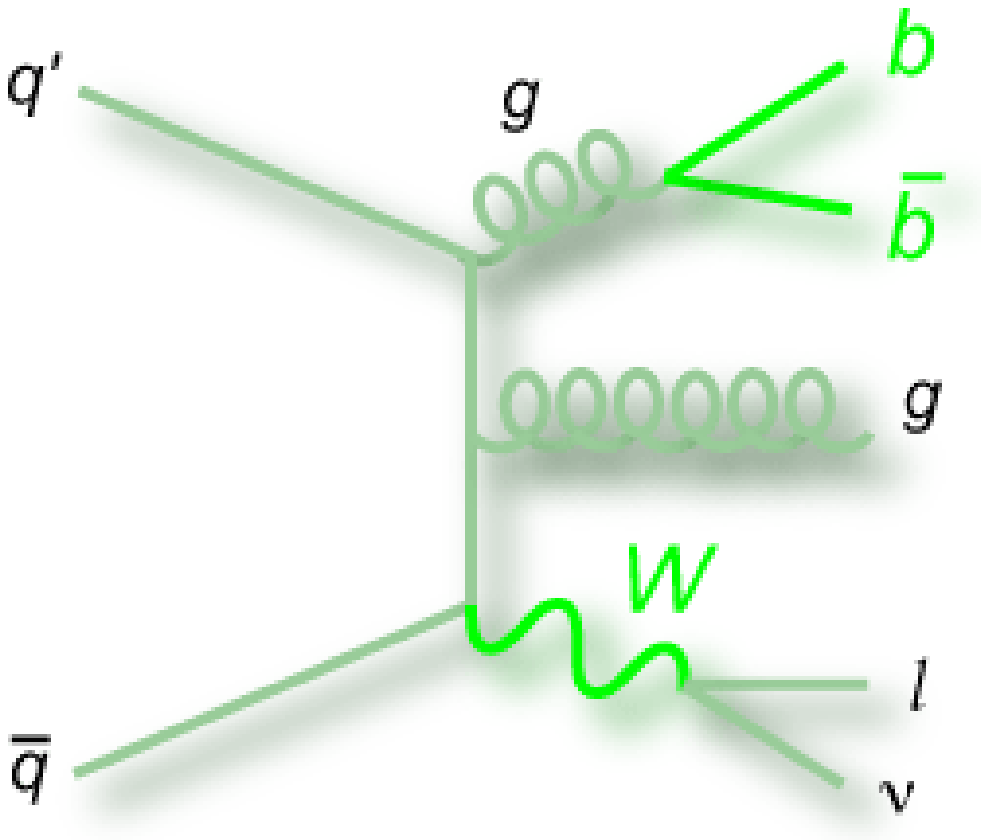}
\hspace{1.6in}
\includegraphics[width=0.25\textwidth]{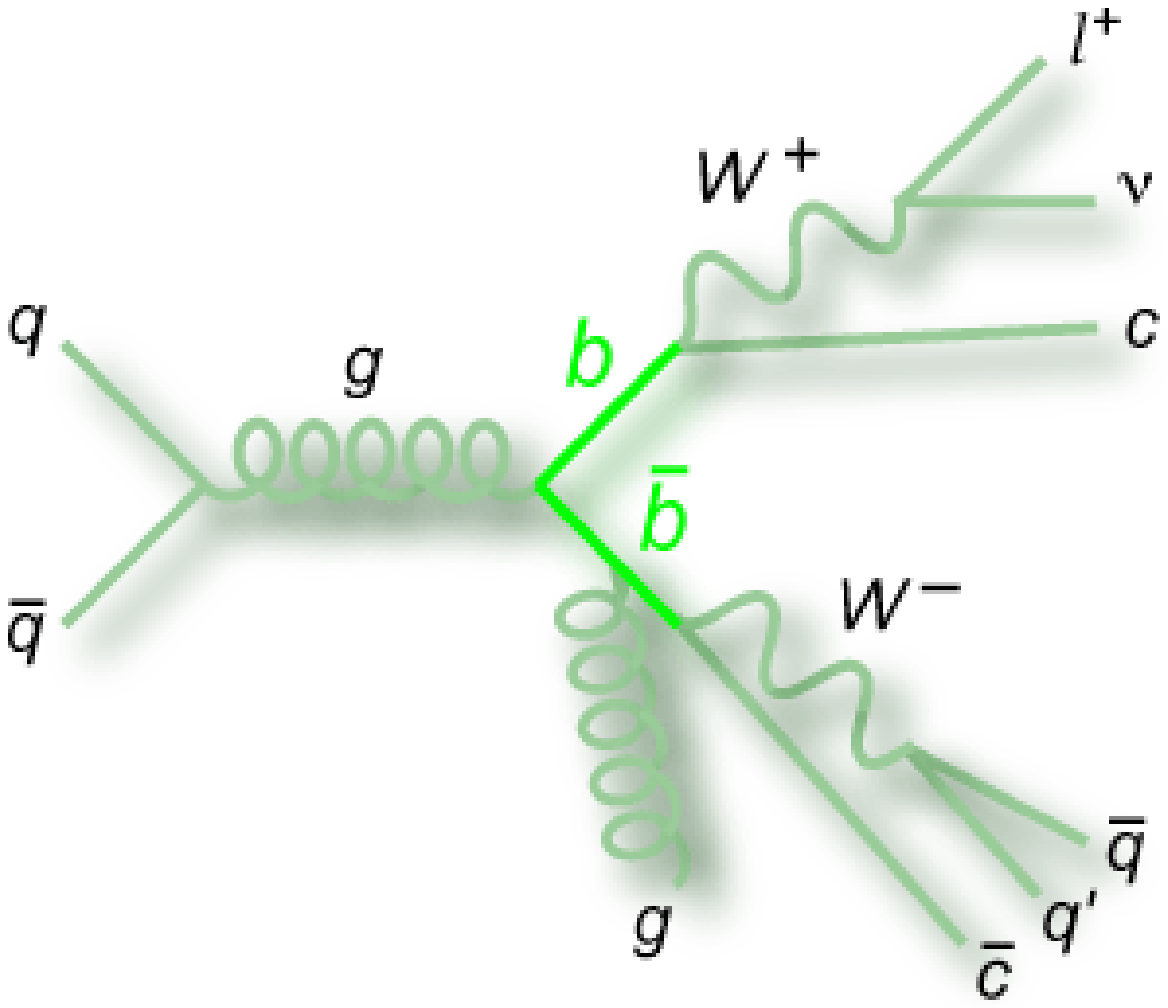}
\hfill
\end{center}
\caption[fig3]{\hspace{0.1in}(a) A tree-level diagram for the
$W{\bbbar}g$ process.\hspace{0.2in} (b) A tree-level diagram for the
${\bbbar}g$ process.}
\label{backgrounds}
\end{figure}

Processes that mimic top quark decays and thus form the backgrounds to
the searches and measurements are dominated for most channels by
events with real $W$ or $Z$ bosons in them ($W$+jets, $Z$+jets, $WW$,
$WZ$, $ZZ$).  A simple $W{\bbbar}g$ three jet event is shown in
Fig.~\ref{backgrounds}(a). The next most common source of background
is from events with misidentified leptons: multijet events with a jet
misidentified as an electron; and {\bbbar}+jets events with a
misidentified electron or muon from a $b$ decay. One example is shown
in Fig.~\ref{backgrounds}(b). There are several additional sources
that can contribute to the backgrounds at a lower level: these include
cosmic rays, multiple {\ppbar} interactions in one bunch crossing, and
pattern recognition mistakes when reconstructing final state
objects. For most {\ttbar} decay channels, processes with a real
$W$~boson and non-$b$ jets with a fake tag are the most difficult to
remove.

\subsection{Bottom Quark Identification}

In order to reject background processes while keeping signal, many top
quark searches and measurements require that at least one jet is
identified as originating from a $b$ quark. CDF and {\dzero} each do
this by using high resolution three-dimensional position information
for the charged tracks near the primary vertex. Fitted tracks are
separated into those from the primary vertex and those from secondary
vertices from the decays of the long-lived $b$~quarks. Several
algorithms are in use, including ones that measure the decay lifetime
significance, and ones that measure the impact parameter
significance. For each experiment, the probability to tag at least one
jet in a {\ttbar} events is about 55\%, with an associated probability
for a fake tag of about 0.4\%.

%---------------------------------------------------------------------
\section{Top Pair Cross Section}

The CDF and {\dzero} collaborations have made a number of measurements
of the {\ttbar} production cross section using the new Run~II
data. These may be compared with the theory value of 6.8~pb, which is
only 0.03\% of the $W$~boson cross section at the Tevatron. All the
main decay modes have been used for these measurements, and the
selected events form the baseline samples for measurements of the top
quark properties. The measurements cannot be combined in a simple
manner to get an improved overall result, since many of them use the
same input data and apply different selection methods to get the final
results.

\subsection{CDF's Measurements}

The CDF collaboration has made 12 measurements of the {\ttbar} cross
section: three using dilepton data; eight with the lepton+jets data;
and one in the alljets channel.

\vspace{0.175in}
\noindent \textbf{Dilepton Measurements}
\vspace{0.075in}

\noindent CDF's first dilepton measurement is a traditional search
where both leptons are identified, no $b$~tagging is used, and the
total transverse energy ($H_T$) of the events is required to be $>
200$~GeV. This yields a measurement with an uncertainty of 44\%. (The
values of the measurements will be given together at the end of this
section.) The second measurement relaxes the identification on one of
the two leptons by just requiring an opposite sign isolated track for
it. This measurement gets a 43\% uncertainty. These two measurements
have been published.~\cite{cdfdilepprl} A third measurement comes from
a simultaneous fit of the cross sections to the data in the
missing-transverse-energy/number-of-jets plane for {\ttbar}, $WW$, and
$Z{\rar}\tau\tau$, plus other processes in a fixed background
component. The resulting cross section has a 32\% uncertainty. Fitting
to one or more variables results in a better quality measurement than
just counting events.

\clearpage
\noindent \textbf{Lepton+Jets Measurements}
\vspace{0.075in}

\noindent CDF have two measurements using lepton+jets data that
require no $b$ identification. The first makes a likelihood fit to the
total transverse energy distribution and get a result with a 51\%
uncertainty. The second measurement combines seven variables in a
neural network and performs a likelihood fit to the network
output. This results in a measurement with a 28\% uncertainty.

CDF have one result where a $b$ jet is identified by the presence of a
muon in the jet. Since the branching fraction for $b$ to $\mu$ is
small, this method has a small acceptance; however, it uses an
independent data sample from the other measurements and can thus
readily be combined with them to get a better overall result. This
measurement has an uncertainty of 77\%.

\begin{figure}[!h!tbp]
\begin{multicols}{2}
\vspace*{0.8in}
\centerline{\protect\includegraphics[scale=0.31]{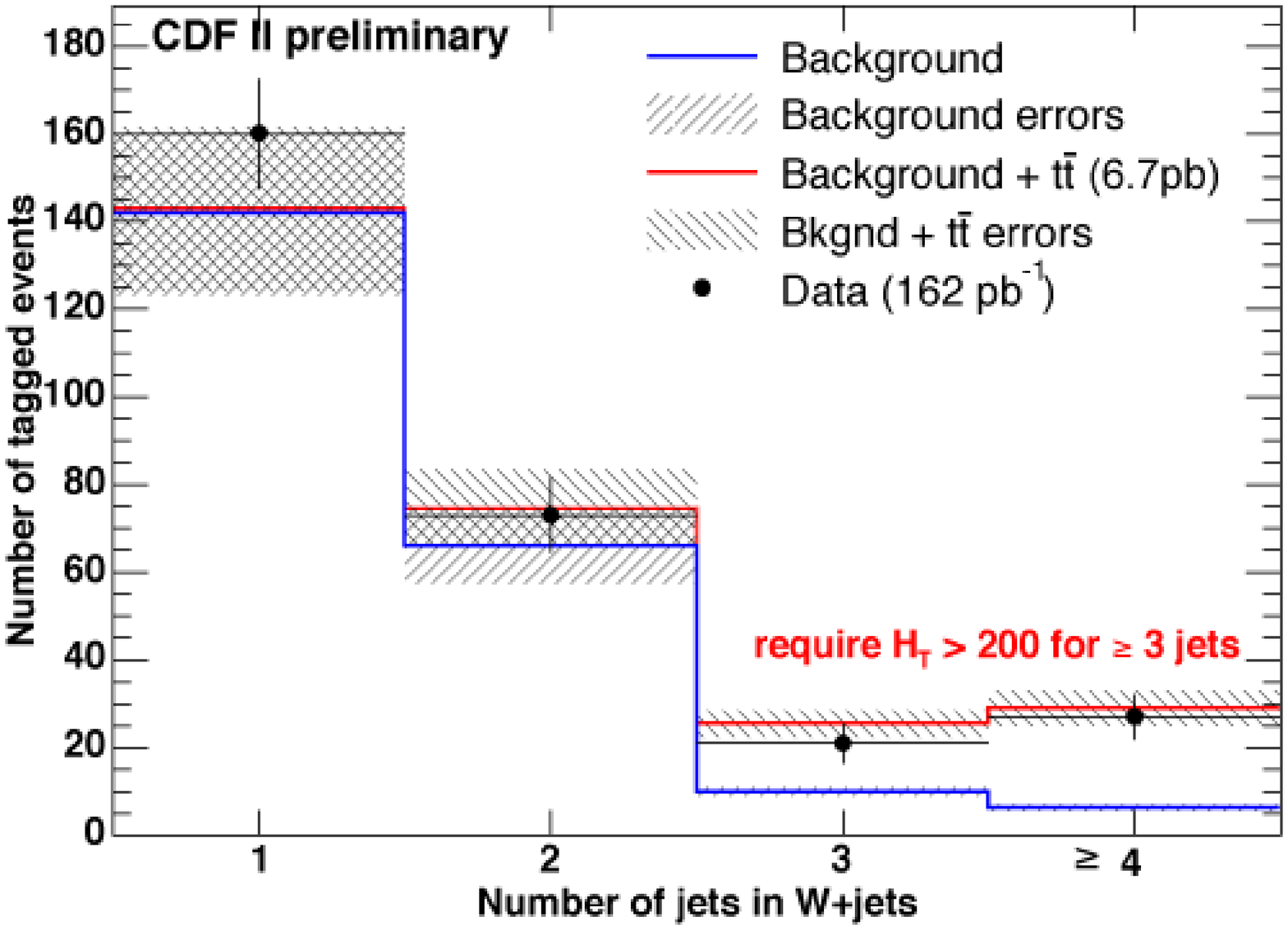}}
\begin{center}
\begin{minipage}{3 in}
\caption[fig4]{\hspace{0.1in}Number of jets after $b$ tagging and
the $H_T$ cut, showing the {\ttbar} excess in the 3- and 4-jet bins.}
\label{cdfljets}
\end{minipage}
\end{center}

\centerline{\protect\includegraphics[scale=0.29]{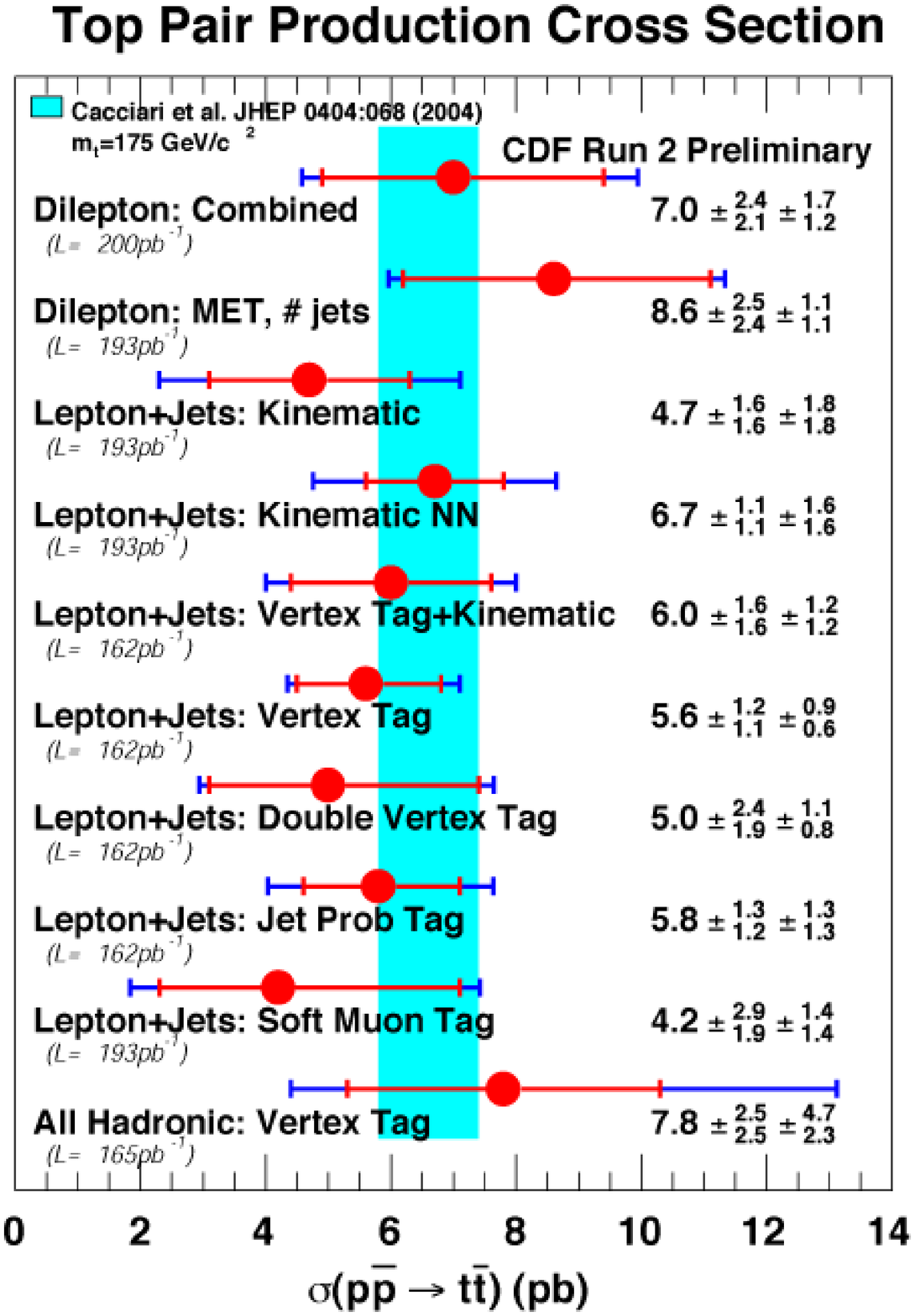}}
\vspace{-0.3 in}
\begin{center}
\begin{minipage}{2.8 in}
\caption[fig5]{\hspace{0.1in}Measurements of the {\ttbar} production
cross section by the CDF collaboration.}
\label{cdftopxsec}
\end{minipage}
\end{center}
\end{multicols}
\end{figure}

CDF have several measurements that use jet and track properties to
identify $b$~jets. The first uses a jet-probability $b$-tagging method
that results in a cross section measurement with a 32\%
uncertainty. The second uses a secondary-vertex $b$-tagging method
with an $H_T > 200$~GeV cut to make a 27\%
measurement,~\cite{cdfljetsb} which is the best result so far in any
channel. They also use this $b$~tagging method with a likelihood fit
to the leading jet transverse energy instead of the $H_T$ cut for a
33\% measurement.~\cite{cdfljetsbfit} They have two double-tagged
measurements: the first applies their standard secondary-vertex
algorithm and obtains a 53\% result;~\cite{cdfljetsb} the second uses
a looser algorithm with higher efficiency and higher fake rate to make
a 37\% measurement. The secondary-vertex result with the $H_T$ cut is
shown in Fig.~\ref{cdfljets}.

\vspace{-0.2in}
\setlength{\columnseprule}{0pt}

\vspace{-0.2in}
\noindent \textbf{Alljets Measurement}
\vspace{0.075in}

\noindent CDF have measured the {\ttbar} cross section in the
all-hadronic decay channel. They cut on four kinematic variables
chosen to reject the multijet background, and require a secondary
vertex $b$ tag. This results in a measurement with a 68\%
uncertainty. Again, although this measurement does not have much
sensitivity on its own, it is independent of those in other decay
channels, and so can be easily combined with them to improve the
overall result. Such combination has not yet been done.

\vspace{0.175in}
\noindent \textbf{CDF {\ttbar} Cross Section Summary}
\vspace{0.075in}

\noindent Figure~\ref{cdftopxsec} shows almost all the results
described above. (Only the loose double $b$ tag is not shown: that
measurement used 162~pb$^{-1}$ of data and found
$\sigma({\ppbar}{\rar}{\ttbar}) = 8.2^{+2.4}_{-2.1}({\rm
stat})^{+1.8}_{-1.0}({\rm syst})$.) The measurements are all
consistent with the NNLO theory band, shown in cyan on the plot.

\subsection{{\dzero}'s Measurements}

The {\dzero} collaboration has made 13 measurements of the {\ttbar}
cross section: four in the dilepton channels; eight with the
lepton+jets data; and one in the alljets channel.

\vspace{0.175in}
\noindent \textbf{Dilepton Measurements}
\vspace{0.075in}

\noindent {\dzero}'s first three dilepton measurements are in the
$ee$, $e\mu$, and $\mu\mu$ channels. Two opposite-sign leptons are
reconstructed, and cuts made on the invariant mass of the dilepton
pair and on the total transverse energy of the event. The uncertainty
on the combined measurement is 41\%.

A novel measurement has been made in the $e\mu$ channel where a
reconstructed secondary vertex in a jet has been used to identify the
presence of a $b$~jet. This method yields a very high purity sample,
and a cross section uncertainty of 54\%, which is statistics dominated
and will improve with a looser tagging definition and more data.

\vspace{0.175in}
\noindent \textbf{Lepton+Jets Measurements}
\vspace{0.075in}

\begin{figure}[!h!tbp]
\begin{multicols}{2}
\vspace*{0.05 in}
\centerline{\protect\includegraphics[scale=0.58]{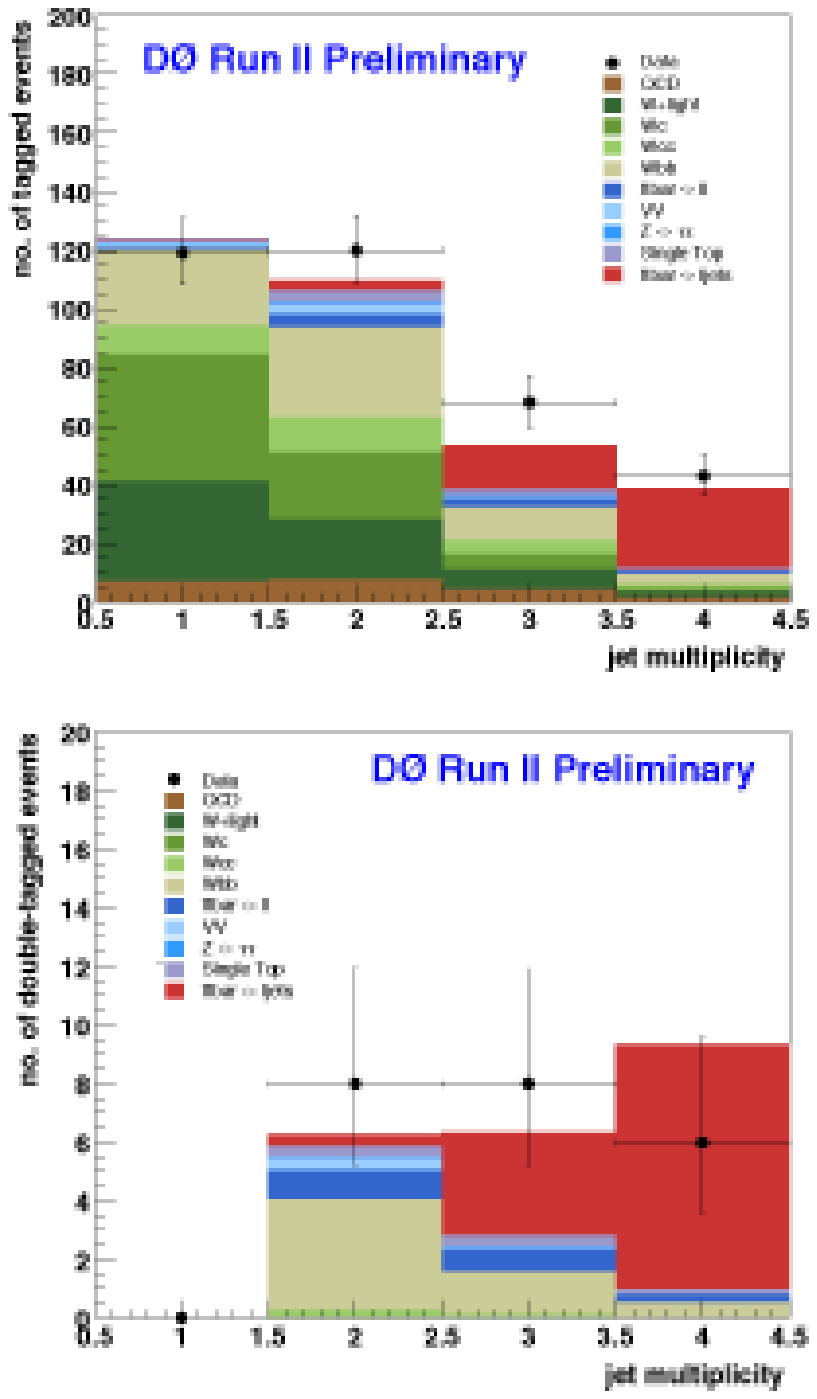}}
\begin{center}
\begin{minipage}{2.8 in}
\vspace{0.1 in}
\caption[fig6]{\hspace{0.1 in}Jet multiplicity for lepton+jets events
with exactly one tagged jet, and with at least two tagged jets.}
\label{d0njets}
\end{minipage}
\end{center}

\vspace*{0.05in}
\centerline{\protect\includegraphics[scale=0.58]{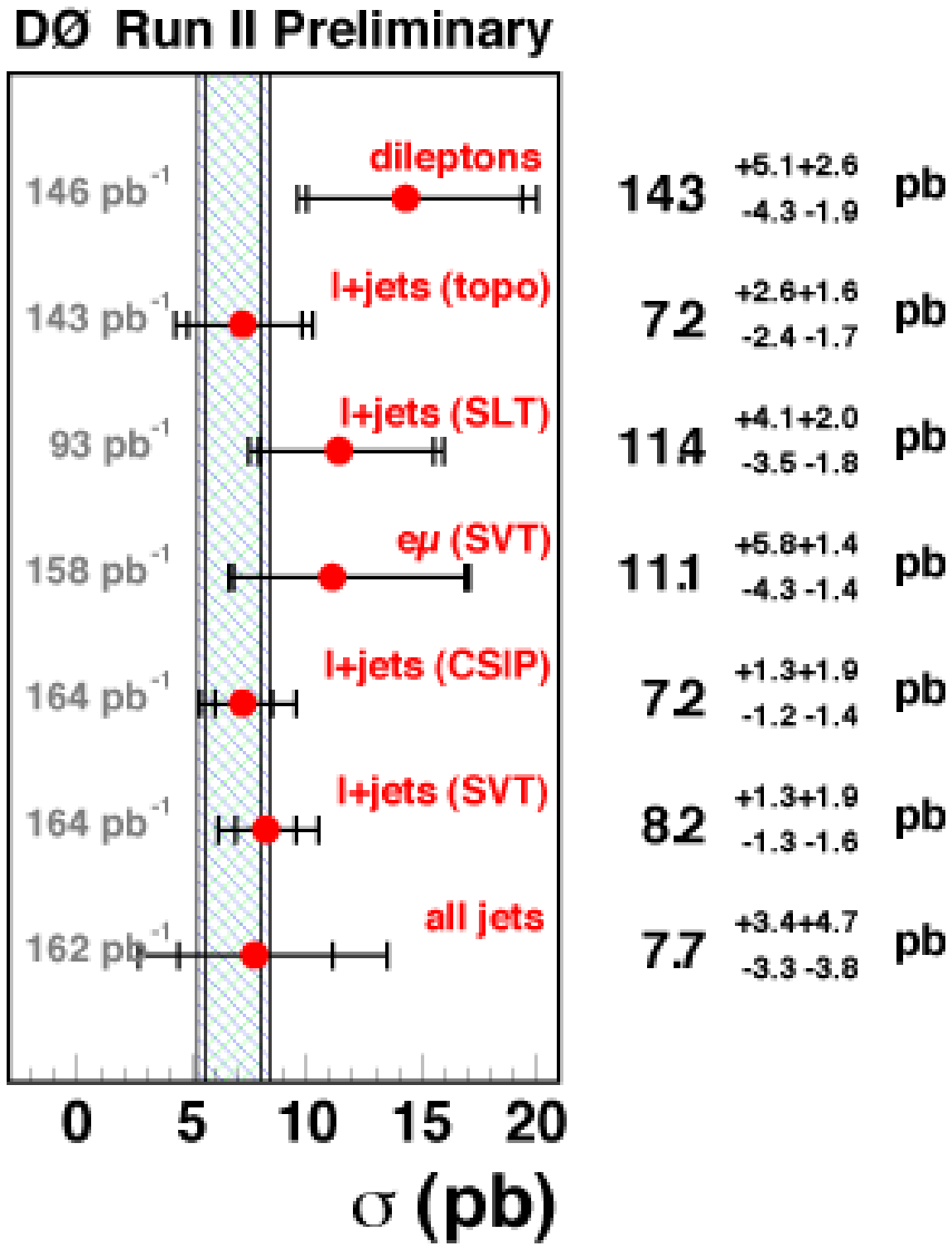}}
\begin{center}
\begin{minipage}{2.8 in}
\vspace{-0.1 in}
\caption[fig7]{\hspace{0.1 in}Measurements of the {\ttbar} production
cross section by the {\dzero} collaboration.}
\label{d0topxsec}
\end{minipage}
\end{center}
\end{multicols}
\end{figure}

\noindent {\dzero} have measured the {\ttbar} cross section in the
single-electron and single-muon decay channels with an untagged
analysis and by using three different $b$-identification methods. The
untagged measurement performs a binned likelihood fit to a likelihood
discriminant variable, and obtains an uncertainty of 42\% for the
combined electron and muon measurements. The first $b$-tagging result
uses the muon-in-jet method, with cuts on aplanarity and $H_T$; this
gives a 41\% uncertainty. The second method identifies $b$~jets with a
secondary vertex, and the third finds $b$~jets with an
impact-parameter significance check. These two methods both make a
likelihood fit for events with exactly three jets, or four or more
jets, and for exactly one tagged jets, or two or more tags. The
resulting cross-section measurements have uncertainties of 29\% for
the secondary-vertex algorithm and 33\% for the impact-parameter
algorithm. The secondary-vertex tagged data are shown in
Fig.~\ref{d0njets}.

\setlength{\columnseprule}{0pt}

\vfill
\clearpage

\vspace{0.175in}
\noindent \textbf{Alljets Measurements}
\vspace{0.075in}

\noindent {\dzero} have used events with at least six jets, including
one with a secondary vertex $b$-identifying tag, to measure the
{\ttbar} cross section in the all-hadronic decay mode. Nine different
kinematic quantities are combined using three sequential neural
networks, and events are counted after cutting on the outputs. The
resulting measurement has a 76\% uncertainty.

\vspace{0.175in}
\noindent \textbf{{\dzero} {\ttbar} Cross Section Summary}
\vspace{0.075in}

\noindent Figure~\ref{d0topxsec} shows the results described above.
The measurements were made with between 92~pb$^{-1}$ and 162~pb$^{-1}$
of data, and are consistent with the NNLO theory band, shown in the
cyan striped band on the plot.

\subsection{Cross Section Summary}

The cross section results are still statistics limited and are
expected to improve significantly in precision as the datasets are
enlarged. The uncertainty on Run~I results was 26--30\%; the best
measurements so far have the same precision already, 27\% (CDF) and
29\% ({\dzero}). With 2~fb$^{-1}$ of data, the uncertainty is expected
to improve to about 10\%, when it will be dominated by the uncertainty
on the integrated luminosity.

%---------------------------------------------------------------------
\section{Single Top Quarks}

\begin{figure}[!h!tbp]
\begin{center}
\includegraphics[width=0.70\textwidth]{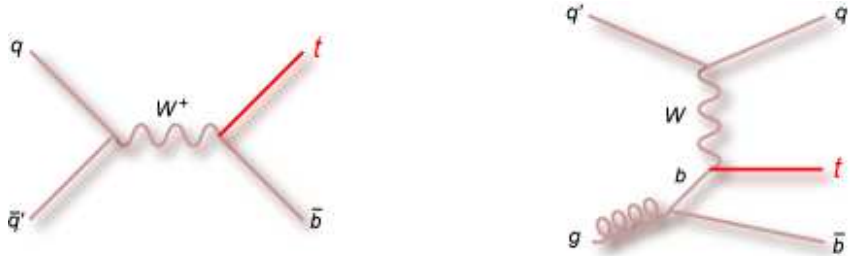}
\end{center}
\vspace{-0.1in}
\caption[fig8]{The main tree-level Feynman diagrams for single top
quark production at the Tevatron Collider.}
\label{feynmansintop}
\end{figure}

In addition to the pair production of top quarks described above, they
are also expected to be produced at the Tevatron singly. There are
three modes, ${\ppbar}{\rar}t{\bbar}$, ${\rar}tq{\bbar}$, and
${\rar}tW$. The third one has a tiny cross section and will be very
difficult to separate from {\ttbar} production. The CDF and {\dzero}
collaborations have both produced results from searches for the first
two modes, known as $s$-channel and $t$-channel
production. Figure~\ref{feynmansintop} shows the main tree-level
diagrams for these processes. The $s$-channel cross section is
calculated to be $0.88 \pm 0.11$~pb at next-to-leading
order,~\cite{sullivan} and the $t$-channel cross section is $1.98
\pm 0.25$~pb at the same order.~\cite{sullivan}

Since there are fewer jets in single top quark events than in most
{\ttbar} ones, the $W$+jets backgrounds are at least ten times higher,
and this is compounded by the lower signal cross
sections. Consequently, it is much more difficult to separate single
top events from background than {\ttbar} ones. Once single top quark
production is observed, the collaborations expect to be able to use it
to measure the CKM matrix element $|V_{tb}|$ without assuming that
there are only three quark generations or CKM unitarity, and hence
will be able to determine the top quark width. Observation is hoped
for with 1--2~fb$^{-1}$ of data.

\subsection{CDF's Measurements}

CDF have recently submitted for publication the results of searches
for $s$-channel single top, for the $t$-channel, and for a combined
search.~\cite{cdfsintop} The searches start with events with a
$W$~boson reconstructed from an isolated lepton and missing transverse
energy, and exactly two jets. Only events with $140{\rm ~GeV} \le
M_{l\nu b} \le 210$~G are used in the search. A likelihood fit is made
to the distribution of lepton charge times pseudorapidity of the
untagged jet, which yields 95\% confidence level (CL) upper limits of
14~pb on the $s$-channel cross section and 10 pb for the
$t$-channel. A separate likelihood fit to the total transverse energy
of the events gives a 95\% CL upper limit on the cross section for
$s$- and $t$-channel combined of 18~pb.

\subsection{{\dzero}'s Measurements}

{\dzero} have released preliminary results on single top quark
searches. They allow between two and four jets, and do not make an
invariant mass cut. Selecting events with high transverse energy of
the lepton, neutrino, and the first two jets yields a 95\% upper limit
of 19~pb on $s$-channel top quark production, 25~pb on $t$-channel
production, and 23~pb on the combined production.

\subsection{Single Top Summary}

Both collaborations are extending their searches by adding more data,
improving the signal acceptances, background models,
$b$-identification tools, uncertainties, and final selection
methods. More sensitive results are expected soon.

%---------------------------------------------------------------------
\section{Top Quark Mass and Other Properties}

Many interesting measurements of top quark properties will be possible
in the future; however, all require high statistics and are not very
significant yet. The properties that will be studied include $gtt$ and
$Wtb$ couplings in production, spin correlations between the top and
antitop, and the production of top from or with new particles. In the
decay, the charge and width will be measurable, together with the rare
CKM branching fractions, $W$ helicities, gluon radiation, transverse
momentum spectra, and other rare decays. Several preliminary
measurements have been made by CDF and {\dzero}: interested readers
are encouraged to read them on the web pages of the Top Groups of the
two collaborations. Here, we will discuss the status of the top quark
mass measurement, a flagship measurement of the Tevatron collider
program.

The CDF and {\dzero} collaborations have been measuring the top quark
mass for the past ten years. The methods have developed significantly
over this time. Early ones used kinematic fitting and compared the
results to templates for many values of the top quark mass. Later
methods added more variables and used likelihoods. The best methods
now use all the available information in the events, but are extremely
time-consuming to implement and to run the calculations. As the
methods have improved, so have the uncertainties on the
measurements. The first results in 1995 had at least a 7\%
error. Current results using Run~II data have a 5\% uncertainty. The
best Run~I result combining all decay channels from both experiments
and using the best method for one of them has a 2.4\% uncertainty. The
goal for Run~II data is to measure the top quark mass to 1\%, which
translates into a $\pm$1.5--2~GeV error.

The world average measurement~\cite{topmassrunI} of the top quark mass
is:
\begin{eqnarray*}
m_{\rm top} = 178.0 \pm 4.3{\rm ~GeV,}
\end{eqnarray*}
\noindent which is dominated by {\dzero}'s measurement in the
lepton+jets channel using the matrix element
method:~\cite{d0massnature,d0massprl} $180.1 \pm 5.3$~GeV.

%\clearpage
\subsection{CDF's Measurements}

CDF have made three measurements using dilepton events, and three
using lepton+jets events. The first dilepton measurement finds the
best fit mass mass after trying all possible solutions for the $\phi$
angles of the two neutrinos in each event, and compares these masses
to template models. It uses 13 data events where both leptons are
fully identified. The uncertainty on the result is 10.7\%. The second
dilepton measurement uses a similar method on almost the same data
events, but instead of the $\phi$ angles as test variable, it uses the
longitudinal momentum of the {\ttbar} system. The uncertainty on the
result is very similar to the first method, 10.5\%. The third method
uses a neutrino weighting algorithm on 19 events where one lepton has
been identified and the other is assumed from an isolated track. This
method yields a measurement with uncertainty 8.3\%. The uncertainties
on the dilepton mass measurements have a systematic component of
7--9~GeV, about the same size as in the lepton+jets
measurements. However, owing to the small numbers of events, the
statistical component of the uncertainties is much larger.

The first CDF measurement with lepton+jets events utilizes kinematic
fitting and mass templates. The method has been applied to events with
no $b$ ~tag, exactly one tag, and two or more tags, and then the three
measurements have been combined. Separately, the uncertainties on the
three results are not as good as the following ones with more
sophisticated techniques, but optimizing on the datasets independently
improves the sensitivity, and the resulting measurement uncertainty is
4.6\%. The second method is similar to the first, but uses the scalar
sum of the first four jets as a variable, together with the standard
reconstructed top quark mass one, for two-dimensional templates. It is
applied to 33 events that have at least one tagged jet, and the
uncertainty in the result is 5.2\%.

The third method applied to tagged lepton+jets data is called the
dynamical likelihood method. It is similar in concept to {\dzero}'s
matrix element method, in that it uses all the information in each
event, not just one or two variables, and each event is weighted by
its probability to be signal or background, so signal-like events
contribute more to the final result than background-like ones. There
are some differences between the two methods, principally in how the
backgrounds are handled. CDF's top mass measurement using the
dynamical likelihood method has a 4.5\% uncertainty, which is the most
precise result from Run~II so far. It is not as good as the best Run~I
result, which had a 2.9\% uncertainty, because the jet energy scale is
not as well calibrated yet.

\begin{figure}[!h!tbp]
\begin{multicols}{2}
\centerline{\protect\includegraphics[scale=0.31]{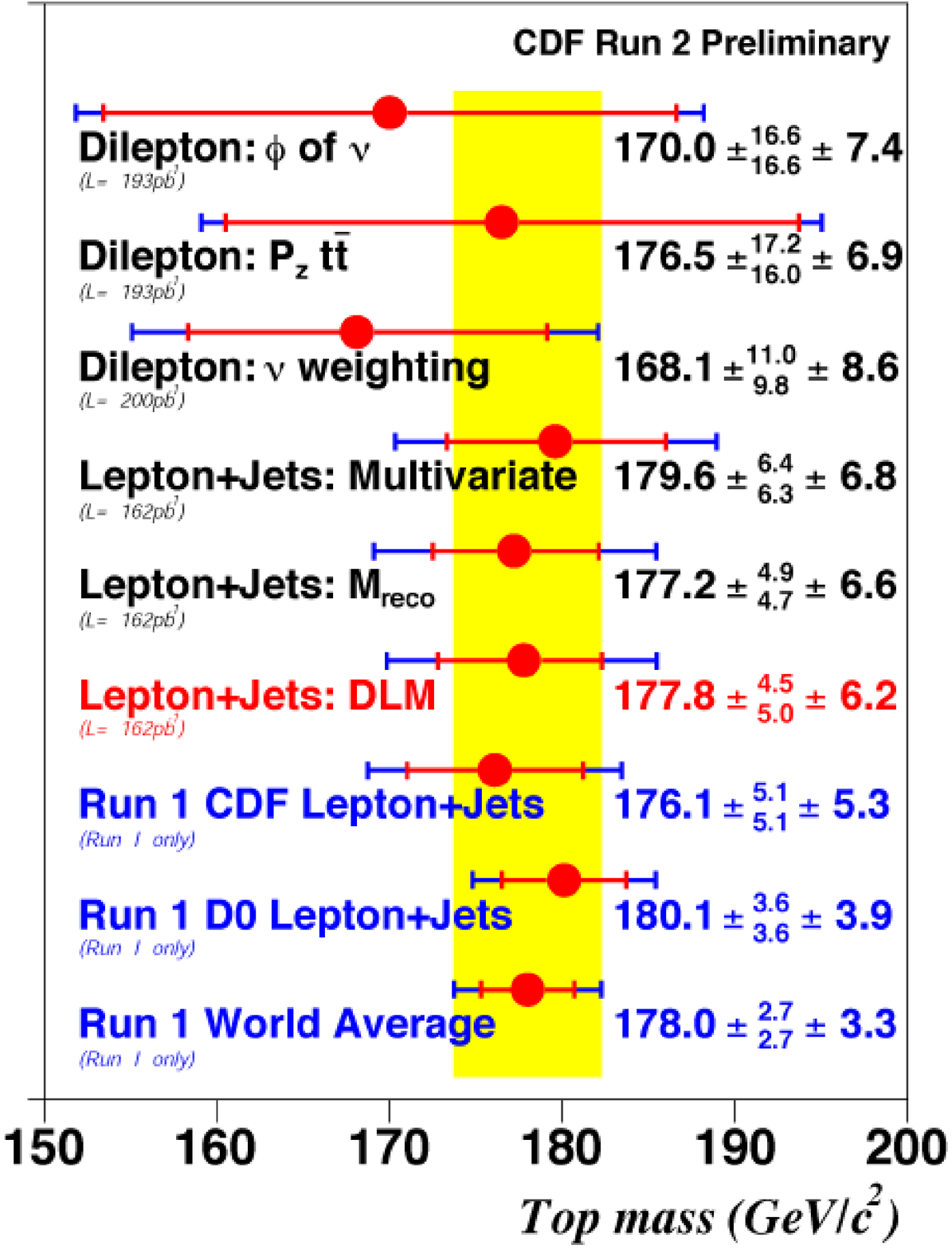}}
\begin{center}
\begin{minipage}{2.8 in}
\vspace{-0.1in}
\caption[fig9]{\hspace{0.1 in}CDF's top quark mass measurements.}
\label{cdftopmass}
\end{minipage}
\end{center}

\vspace*{0.1 in}
\centerline{\protect\includegraphics[scale=0.31]{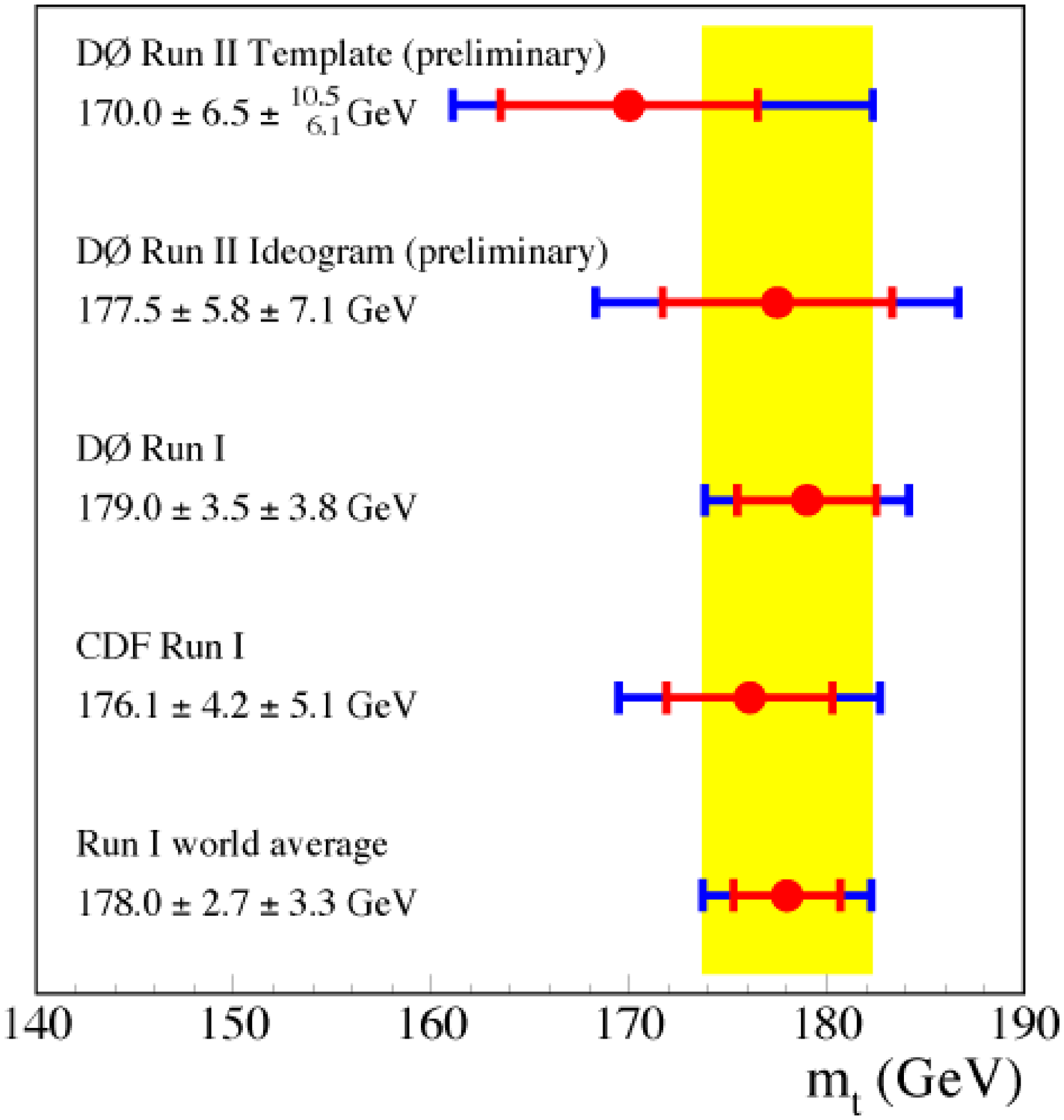}}
\begin{center}
\begin{minipage}{2.8 in}
\vspace{-0.1in}
\caption[fig10]{\hspace{0.1 in}{\dzero}'s top quark mass measurements.}
\label{d0topmass}
\vspace{0.4in}
\end{minipage}
\end{center}
\end{multicols}
\end{figure}

CDF's top quark mass measurements are shown in
Fig.~\ref{cdftopmass}. The yellow band shows the world average
measurement from published Run~I results.

\vspace{0.1in}
\setlength{\columnseprule}{0pt}

\clearpage

\subsection{{\dzero}'s Measurements}

{\dzero} have made two measurements of the top quark mass using Run~II
data, both in the lepton+jets decay channel. The first method uses
templates for signal and background, and fitting to 86 data events
yields a measurements with an uncertainty of 7.1\%. The second method,
called ``ideogram,'' uses an analytical likelihood for each event to
see whether it is signal or background. The resulting mass measurement
has an uncertainty of 5.2\%. As with CDF's measurements, this better
result is from the method that uses more information about each
event. {\dzero}'s top quark mass measurements are shown in
Fig.~\ref{d0topmass}. The yellow band shows the same world average
measurement as in the neighboring CDF plot.

\subsection{Top Mass Summary}

New measurements using Tevatron data from Run~II are now appearing,
and in the future the precision will be better than 2~GeV. Since the
top quark couples strongly to the Higgs boson, and plays a critical
role in loop corrections, one can combine the mass information with
other precision measurements to determine the most favored mass region
for the Higgs boson. The Run~II top quark mass results have not yet
been published or combined with previous measurements; using the
published values instead, the latest best fit Higgs boson mass is
$114_{-45}^{+69}$~GeV, and the 95\% confidence level upper limit on
the Higgs boson mass is 260~GeV.\cite{rentonichep}

%---------------------------------------------------------------------
\section{Future Top Quark Physics}

The future of top quark physics is very bright. Whereas the Tevatron
produces about $10^4$ top quark pairs per year, the Large Hadron
Collider (LHC) will produce $10^7$ pairs per year for the first three
years, and $10^8$ per year after that. This is because both the
energy, at 14~TeV, and the luminosity will be much higher, boosting
both the cross section and the interaction rate. Because the LHC is a
proton-proton machine, and not proton-antiproton, about 90\% of the
rate will come from the $gg$ initial state and the remaining 10\% from
{\qqbar}, almost exactly opposite to the situation at the Tevatron.

At the International Linear Collider (ILC), if the electron-positron
collisions are at the {\ttbar} threshold, about 360~GeV, then $10^6$
pairs will be produced per year. The $e^+e^-{\rar}{\ttbar}$ cross
section is much lower than at the Tevatron, but the ILC luminosity
will be much higher. In addition, the backgrounds will be much lower
than at the Tevatron, and many precision measurements will be
possible.

\begin{figure}[!h!tbp]
\begin{center}
\includegraphics[width=0.37\textwidth]{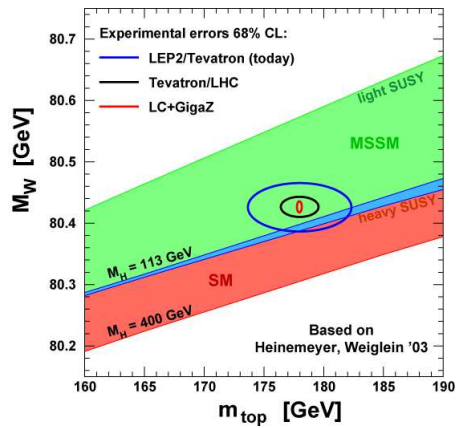}
\end{center}
\begin{center}
\begin{minipage}{4.5in}
\begin{center}
\caption[fig11]{Predicted sensitivity to SM and MSSM Higgs
boson mass from measurements of the top quark mass and $W$~boson mass
at future colliders.\cite{heinemeyer}}
\label{futurelimits}
\end{center}
\end{minipage}
\end{center}
\end{figure}

At the LHC and ILC, all top quark properties, including SM and non-SM
couplings, and rare production and decay modes, will be studied in
detail. To give one example of the expected precision, let us consider
the top quark mass. Similar precision is expected from both the
Tevatron and the LHC, about 1--2~GeV. These measurements will be
limited by our understanding of the final state radiation. At the ILC,
one can scan the {\ttbar} threshold and fit $m_{\rm top}(1S)$,
$\alpha_{\rm s}(M_Z)$, $\Gamma_{\rm top}$, and $g_{tH}$ to
measurements of $\sigma_{\ttbar}$, $p_{\rm top}$, and $A^{\rm
top}_{\rm FB}$, to yield a measurement of $m_{\rm top}(1S)$ to 20~MeV
precision.\cite{martinez} Converting $m_{\rm top}(1S)$ to $m_{\rm
top}(\bar{\rm MS})$ limits the $m_{\rm top}(\bar{\rm MS})$ uncertainty
to about 100~MeV.\cite{hoang} The future precision on the Higgs boson
from measurements of the $W$~boson mass and top quark mass are shown
in Fig.~\ref{futurelimits}.

\vfill
\clearpage

%---------------------------------------------------------------------
\section{Summary}

The Tevatron {\ppbar} Collider is the only top quark factory until the
Large Hadron Collider turns on at the end of 2007. The Tevatron is
meeting performance expectations and the CDF and {\dzero}
collaborations are collecting data at high efficiency. About 80 times
more data in Run~II is expected than was collected in Run~I
(100~pb$^{-1}$ ${\rar}$ 8~fb$^{-1}$). Many first measurements with the
Run~II data are now available; all are consistent with the Standard
Model predictions. More data is needed to reduce both the statistical
and systematic uncertainties on these measurements. In addition, more
time is required to apply more sophisticated analysis methods to the
measurements. As this begins to happen, a precision top quark physics
program will emerge. The top quark will provide a unique window into
hidden parts of the Standard Model and hopefully many regions beyond.

\section*{Acknowledgments}

I would like to thank the organizers and hosts of the Rencontres du
Vietnam in Hanoi for their gracious and thoughtful hospitality. To
travel to Vietnam and meet the generous people and see the beautiful
countryside and fascinating city life was a great opportunity, one I
shall never forget. The meeting itself was first class and extremely
interesting.

%---------------------------------------------------------------------


\section*{References}

\begin{thebibliography}{99}

\bibitem{topmassrunI}
The CDF Collaboration, The {\dzero} Collaboration, and the Tevatron
Electroweak Working Group, ``Combination of CDF and {\dzero} results
on the top-quark mass,'' hep-ex/0404010.

\bibitem{topxsectheory}
N.~Kidonakis and R. Vogt,
``Next-to-next-to-leading order soft-gluon corrections in top quark
hadroproduction,''
{\em Phys. Rev.} D {\bf 68}, 114014 (2003).

\bibitem{cdfdilepprl}
D.~Acosta {\it et al.}, (CDF Collaboration),
``Measurement of the {\ttbar} production cross section in {\ppbar}
collisions at $\sqrt{s} = 1.96$~TeV using dilepton events,''
{\em Phys. Rev. Lett.}  {\bf 93}, 142001 (2004).

\bibitem{cdfljetsb}
D.~Acosta {\it et al.}, (CDF Collaboration),
``Measurement of the {\ttbar} production cross section in {\ppbar}
collisions at $\sqrt{s} = 1.96$~TeV using lepton+jets events with
secondary vertex $b$ tagging,''
submitted to {\em Phys. Rev.} D, hep-ex/0410041.

\bibitem{cdfljetsbfit}
D.~Acosta {\it et al.}, (CDF Collaboration),
``Measurement of the {\ttbar} production cross section in {\ppbar}
collisions at $\sqrt{s} = 1.96$~TeV using kinematics fitting of
$b$-tagged lepton+jets events,''
submitted to {\em Phys. Rev.} D, hep-ex/0409029.

\bibitem{sullivan}
B.W.~Harris, E.~Laenen, L.~Phaf, Z.~Sullivan, and S. Weinzierl,
``Fully differential single-top-quark cross section in next-to-leading
order QCD,''
{\em Phys. Rev.} D {\bf 66}, 054024 (2002).\\
Z.~Sullivan,
``Understanding single-top-quark production and jets at hadron
colliders,''
to appear in {\em Phys. Rev.} D, hep-ph/0408049.

\bibitem{cdfsintop}
D.~Acosta {\it et al.}, (CDF Collaboration),
``Search for electroweak single top quark production in {\ppbar}
collisions at $\sqrt{s} = 1.96$~TeV,''
submitted to {\em Phys. Rev. Lett.}, hep-ex/0410058.

\bibitem{d0massnature}
V.M.~Abazov {\it et al.}, ({\dzero} Collaboration),
``A precision measurement of the mass of the top quark,''
{\em Nature} {\bf 429}, 638 (2004).

\bibitem{d0massprl}
V.M.~Abazov {\it et al.}, ({\dzero} Collaboration),
``New measurement of the top quark mass in lepton+jets {\ttbar}
events at {\dzero},''
submitted to {\em Phys. Rev. Lett.}, hep-ex/0407005.

\bibitem{rentonichep}
P.B.~Renton,
``Electroweak fits and constraints on the Higgs mass,''
submitted to the 32nd International Conference on High Energy Physics,
Beijing, China, (August 2004), hep-ex/0410177.

\bibitem{martinez}
M.~Martinez and R.~Miquel,
``Multiparameter fits to the {\ttbar} threshold observables at a
future $e^+e^-$ linear collider,''
{\em Eur. Phys. J.} {\bf C 27}, 49 (2003).

\bibitem{hoang}
A.H.~Hoang, A.V.~Manohar, I.W.~Stewart, and T. Teubner,
``Threshold {\ttbar} cross section at next-to-next-to-leading
logarithmic order,''
{\em Phys. Rev. D} {\bf 65}, 014014 (2002).

\bibitem{heinemeyer}
S.~Heinemeyer and G.~Weiglein,
``The MSSM in the light of precision data,''
in the Proceedings of the Mini-Workshop on Electroweak Precision Data
and the Higgs Mass, Zeuthen, Germany, (February 2003), hep-ph/0307177.

\end{thebibliography}
\end{document}